\documentclass[amsmath,11pt]{article}

\usepackage{amssymb,amssymb,epsfig,bm}
\usepackage{caption}
\usepackage{subcaption}

\date{\empty}

\begin{document}
\title{\bf General relativity, early galaxy formation\\ and the JWST observations\footnote{Essay received ``Honorable Mention'' in the 2025 Gravity Research Foundation (GRF) essay competition.}}

\author{Christos G. Tsagas${}^{1,2}$\\ {\small ${}^1$Section of Astrophysics, Astronomy and Mechanics, Department of Physics}\\ {\small Aristotle University of Thessaloniki, Thessaloniki 54124, Greece}\\ {\small ${}^2$Clare Hall, University of Cambridge, Herschel Road, Cambridge CB3 9AL, UK}\\ {\small (email address: tsagas@astro.auth.gr)}}

\maketitle

\begin{abstract}
The James Webb Space Telescope has recently detected massive, fully formed, galaxies at redshifts corresponding to few hundred million years after the Big-Bang. However, our current cosmological model cannot produce such massive systems so early in the lifetime of the universe. A number of theoretical solutions have been proposed, but they all appeal to exotic new physics and introduce rather excessive fine-tuning. In this essay, we outline a theoretical answer to the early galaxy-formation question, which operates within standard general relativity and standard cosmology, without appealing to any new physics. Instead, we account for the effect of a well established feature of our universe. This feature, which has so far been kept in the margins of mainstream cosmology, are the peculiar velocities.
\end{abstract}

In the last years, there have been several claims that the James Webb Space Telescope (JWST) has detected massive galaxies at redshifts much higher than expected (e.g.~see~\cite{Cetal}). In fact, early indications of unusually high-redshift galaxies were already given by the Hubble Space Telescope (HST) nearly a decade ago~\cite{Metal}. More recently, in January 2024, a massive galaxy (JADES-GS-z14-0) of hundreds of millions Solar masses was spectroscopically confirmed at a redshift of 14.32. The extent of the source implied that its light was coming mainly from young stars, rather than from emission near a supermassive black hole. In addition, the detection of oxygen emission lines suggested that multiple generations of very massive stars had already existed before the galaxy was detected. Given that relatively small regions of the sky have been searched so far, it is  likely that more such luminous galaxies will be found by JWST in the years to come, perhaps at even higher redshifts. The question then is: \textit{What physical mechanism can form such bright, massive galaxies within only few hundred million years from the Big Bang?}

The possibility that the early universe, at an age much younger than one billion years, was densely populated by massive, well developed galaxies appears to be at direct odds with the current standard cosmological model. Indeed, within the limits of the $\Lambda$CDM paradigm, high-redshift galaxies like those identified in the JWST data should not had the time to form (e.g.~see~\cite{Hetal}). Several theoretical solutions have been proposed to resolve the emerging \textit{early growth problem} and the popular approach is to (somehow) enhance the (actual or effective) gravitational field. By so doing, one could expedite the process of structure formation and thus allow massive galaxies to form at earlier cosmological times. So far, however, even the most conservative of the proposed theoretical solutions either introduce ad hoc modifications to general relativity, or add new free parameters to the existing structure formation models (or both). Moreover, all these scenarios involve considerable fine tuning (see~\cite{LB} for a representative though incomplete list). This essay takes a different approach to the question of early galaxy formation. Instead of appealing to exotic new physics, we will consider the possibility that we have not yet exhausted the existing theories. Put another way, we will demonstrate that there are still unexplored features in our current gravitational theory that could provide the answer. In our case, the unexplored feature are the ubiquitous peculiar motions and their (as yet unexplored) purely relativistic contribution to the gravitational field.

To begin with, let us recall the standard structure-formation scenario, which entirely bypasses peculiar velocities. According to this model, the post-recombination evolution of linear density perturbations in the pressure-free matter (baryonic or not) is governed by the differential equation (e.g.~see~\cite{Pa,TCM})
\begin{equation}
\ddot{\delta}+ 2H\dot{\delta}- {3\over2}\,H^2\delta= 0\,, \label{lDelddot}
\end{equation}
Here, $\delta=\delta\rho/\rho$ represents the familiar density contrast (with $\rho$ being the matter density) and $H=\dot{a}/a$ is the Hubble parameter, with $a=a(t)$ being the cosmological scale factor. Since $a\propto t^{2/3}$ and $H=2/3t$ after matter-radiation equality, the above accepts the power-law solution
\begin{equation}
\delta= \mathcal{C}_1t^{-1}+ \mathcal{C}_2t^{2/3}\,,  \label{lDel}
\end{equation}
where $\mathcal{C}_{1,2}$ are the integration constants. Therefore, linear perturbations in the density of the pressureless matter grow as $\delta\propto t^{2/3}\propto a$, once the universe has entered its dust era. There is a serious problem with this solution, however, because the growth rate is too slow. Indeed, in line with the Cosmic Microwave Background (CMB) data, $\delta\simeq10^{-5}$ at the time of recombination, namely at $z\simeq10^3$. Then, the current value of the density contrast today should be $\delta_0\simeq10^{-2}$. This lies too far from the nonlinear threshold of $\delta=1$, where structure formation is expected to begin in earnest. The problem is typically solved by introducing Cold Dark Matter (CDM), an as yet undetected species that (in contrast to baryons) can start agglomerating gravitationally well before equipartition. As a result, by the time the universe enters its dust epoch, there are ``gravitational wells'' of CDM, which can accelerate the growth of baryonic density perturbations and in so doing expedite the whole process of structure and galaxy formation. Although CDM plays a key role in the current cosmological model (the $\Lambda$CDM paradigm), it appears incapable of accommodating massive, fully formed, galaxies at the high redshifts reported by JWST.

Modifying the form of differential equation (\ref{lDelddot}) can change the evolution of the density contrast and potentially trigger a faster linear growth for it. It could then be theoretically feasible to accelerate structure formation and form massive galaxies considerably earlier than it is currently expected and, in so doing, explain those found in the JWST data. So far, all this can be achieved by appealing to some sort of new physics, together with the considerable fine-tuning of the newly introduced free parameters. In this essay, we will discuss a way of modifying Eq.~(\ref{lDelddot}) within the realm of general relativity, without breaking away from the Friedmann cosmological models and without introducing any other new physics. Instead, we will account for the effects of an observationally established feature of the universe, which however has been left at the margin of mainstream cosmological research. This neglected feature are the peculiar velocities.

Bulk peculiar motions are commonplace in the universe, confirmed by numerous surveys (e.g.~see~\cite{WFH}). So, it is fair to say that no real observer follows the mean universal expansion, but we all have some finite peculiar velocity with respect to it. Our galaxy and the Local Group, for example, move at approximately 600~km/sec with respect to the coordinate system of the Cosmic Microwave Background (CMB), which defines the rest-frame of the cosmic expansion. These bulk flows are believed to have started as weak velocity perturbations around the time of equipartition/recombination, which have grown to the observed large-scale bulk flows by the ever increasing inhomogeneity of the post-recombination universe.

Peculiar flows are nothing else but matter in motion and moving matter implies nonzero energy flux. At the linear perturbative level, this \textit{peculiar flux} is given by $q_a=\rho v_a$, with $v_a$ representing the peculiar velocity (e.g.~see~\cite{TCM}). In general relativity, as opposed to Newtonian gravity, energy fluxes ``gravitate'' as well, since they also contribute to the energy-momentum tensor of the matter. Although, the (purely relativistic) gravitational input of the peculiar flux is undisputed, its implications have been typically bypassed. Among others, the peculiar-flux contribution to the relativistic gravitational field feeds into the conservation laws and eventually emerges into the linear equations governing the evolution of density perturbations. As a result, when peculiar velocities are accounted for, Eq.~(\ref{lDelddot}) is replaced by
\begin{equation}
\ddot{\delta}+ 2H\dot{\delta}- {3\over2}\,H^2\delta= -3aH\ddot{v}- 3aH^2\dot{v}+ {9\over2}\,aH^3v\,,  \label{ltDelddot}
\end{equation}
with $v$ representing the magnitude of the peculiar-velocity field (see~\cite{TCM} for the initial nonlinear expression of (\ref{ltDelddot})). Comparing the above to differential equation (\ref{lDelddot}), one immediately notices the difference in the profiles of the two linear formulae. The physical reason behind this difference is the aforementioned (purely relativistic) gravitational contribution of the peculiar flux.

In Newtonian theory linear peculiar velocities grow as $v\propto t^{1/3}$ (e.g.~see~\cite{Pe}). Substituting this rate into the right-hand side of (\ref{ltDelddot}), while keeping in mind that $a\propto t^{2/3}$ and $H=2/3t$ in the dust epoch, leads to the familiar growth rate of $\delta\propto t^{2/3}$ for the density contrast (see solution (\ref{lDel}) above). The relativistic treatment of linear peculiar velocities, on the other hand, supports a stronger linear growth with $v\propto t^{4/3}$~\cite{TT}. Again, the reason for the difference between the Newtonian and the relativistic results is the peculiar-flux input to the gravitational field. Inserting the relativistic growth rate into the right-hand side of (\ref{ltDelddot}), the latter accepts the solution
\begin{equation}
\delta= \mathcal{C}_1t^{-1}+ \mathcal{C}_2t^{2/3}+ \mathcal{C}_3t\,,  \label{ltDel}
\end{equation}
where $\mathcal{C}_{1,2,3}$ are constants. The first two modes in the above reproduce the standard solution (see (\ref{lDel}) earlier), while the third is a new mode conveying the peculiar-velocity effect. Therefore, when the latter is included, the linear growth-rate of the density contrast increases from $\delta\propto t^{2/3}$ to $\delta\propto t$. This raises the current value of the density contrast from $\delta_0\simeq10^{-2}$ to $\delta_0\simeq10^{-1/2}$, which lies tantalising close to the nonlinear threshold.

Therefore, either by mimicking the gravitational effects of the CDM, or by acting together with it, peculiar-velocity perturbations can enhance the standard linear growth of density inhomogeneities after recombination. Technically speaking, the aforementioned enhancement results from the peculiar-flux contribution to the energy-momentum tensor, which then feeds into the relativistic conservations laws and eventually emerges into the linear differential equations monitoring the evolution of density perturbations. Physically speaking, it is the input of the peculiar energy-flux that enhances the relativistic gravitational field and leads to a stronger linear growth-rate for the density perturbations. This in turn expedites the process of structure formation and in so doing it can produce massive galaxies at higher redshifts, perhaps like those detected by the JWST. Crucially, all this is achieved without modifying general relativity, or by introducing any new physics, but by simply accounting for the fundamentally different way Newton's and Einstein's theories treat both the gravitational field and its sources. In a sense, one could say that, within the framework of general relativity, \textit{peculiar flows
gravitate}.

In summary, linear structure formation proceeds more efficiently in the presence of peculiar motions, than in their absence. This makes it more likely to produce massive galaxies at high redshifts, like those recently found in the JWST data, when the effect of peculiar velocities is properly accounted for. What distinguishes this scenario from the rest of the proposed solutions to the early galaxy-formation problem,  is that everything happens within the realm of general relativity and standard cosmology and without introducing any exotic new physics. Instead of appealing to speculative and often far-fetched scenarios, one simply needs to account for the effects of peculiar motions. The widespread presence of the latter is a hard observational fact and perhaps it is about time to fully incorporate their study into the mainstream cosmological research and into our structure-formation models. After all, having entered the era of precision cosmology, we cannot go on ignoring peculiar velocities indefinitely.\\

\textbf{Acknowledgements:} This work was supported by the Hellenic Foundation for Research and Innovation (H.F.R.I.), under the ``First Call for H.F.R.I. Research Projects to support Faculty members and Researchers and the procurement of high-cost research equipment Grant'' (Project Number: 789).\\

\end{document}